# Solving the Schrödinger Equation for a Charged Particle in a Magnetic Field using the Finite Difference Time Domain Method

**I Wayan Sudiarta [1, *] and D J Wallace Geldart [1, 2]**

[1] Department of Physics and Atmospheric Science, Dalhousie University, Halifax, NS  B3H 3J5 Canada

[2] School of Physics, University of New South Wales, Sydney, NSW 2052, Australia

E-mail : sudiarta@dal.ca

**Abstract**

We extend our finite difference time domain method for numerical solution of the Schrödinger equation to cases where eigenfunctions are complex-valued. Illustrative numerical results for an electron in two dimensions, subject to a confining potential $V(x,y)$, in a constant perpendicular magnetic field demonstrate the accuracy of the method.

## 1. Introduction

In a recent paper [1] we used the finite difference time domain (FDTD) method to solve numerically the Schrödinger equation for the ground state and excited state eigenvalues and eigenfunctions for a variety of typical examples of a single particle in one, two, and three-dimensional potential wells. To apply the FDTD method, the Schrödinger equation

$$i\hbar \frac{\partial}{\partial t}\psi(\mathbf{r},t) = \hat{H}\psi(\mathbf{r},t) \qquad (1)$$

is first transformed into a diffusion equation in imaginary time $\tau = it/\hbar$,

$$\frac{\partial}{\partial \tau}\psi(\mathbf{r},\tau) = -\hat{H}\psi(\mathbf{r},\tau). \qquad (2)$$

The resulting differential equation is then numerically solved using the FDTD method. Note that Eq. (2) is strictly analogous to a diffusion equation, with a real-valued wave function $\psi(\mathbf{r},\tau)$ as solution, if the Hamiltonian $\hat{H}$ is real-valued (no explicit imaginary terms with factors of $i$) and Hermitian (with suitable boundary conditions). This was the case for all of the examples which were considered in [1].

---

* Corresponding Author. E-mail: sudiarta@dal.ca

The purpose of this paper is to extend the FDTD method to situations where the Hamiltonian has explicit imaginary terms (so is complex-valued) but still remains Hermitian. This occurs in a number of applications, in particular where time reversal symmetry is broken as in the case of charged particles in magnetic fields. In general, the solution of Eq. (2) then yields a complex wavefunction $\psi(\mathbf{r},\tau)$. Since the Hamiltonian is Hermitian, the eigenvalues are still real even though the wave functions are complex.

In this paper we focus attention on single-electron systems in two dimensions in a constant external magnetic field. This is sufficient to illustrate the procedure for practical applications to quasi-two-dimensional nanostructure devices. The 2D character arises when the effective potential generated at the interface between semiconductor and insulator regions confines the motion of electrons to the plane of the interface. We take the motion of the electron to be subject to a planar confining potential $V(x,y)$ and the magnetic field to be perpendicular to the 2D plane. This system represents a single-electron 2D quantum dot in a magnetic field.

To demonstrate the applicability of the FDTD method to such systems two examples of $V(x,y)$ will be considered; (i) an isotropic harmonic oscillator potential and (ii) an anisotropic quartic potential. In the first case, an exact analytical solution is known so the accuracy of the FDTD results can be tested and directly verified. In the second case, no exact results are available for general anisotropic anharmonic confining potentials. However the FDTD method is equally efficient in cases where the confining potential well has no particular simplifying symmetry properties. This is a very important practical feature of the FDTD algorithm. For this reason we also apply the FDTD method to an anisotropic quartic confining potential. Results for energy eigenvalues and complex eigenfunctions are obtained and their accuracy is verified by an independent matrix diagonalization procedure.

For these illustrative models for a quantum dot in a magnetic field the Hamiltonian is

$$\hat{H} = \frac{1}{2m}\left[-i\hbar\nabla + e\mathbf{A}(x,y)\right]^2 + V(x,y) - \boldsymbol{\mu}\cdot\mathbf{B} \qquad (3)$$

where $m, -e, \boldsymbol{\mu}$ denote the electron mass, charge and magnetic moment. In general $\psi(\mathbf{r},\tau)$ with $\mathbf{r} = (x,y)$ in Eq. (2) is a spinor with components $\psi_\sigma(\mathbf{r},\tau)$, $\sigma = \pm 1$. Since the magnetic field $\mathbf{B}$ is constant, the spin and space variables are not coupled. The Zeeman energy in Eq. (3) contributes $\pm \mu B$ to eigenvalues but does not affect the space dependence of $\psi_\sigma(\mathbf{r},\tau)$. To demonstrate the FDTD method for determination of wave functions, it is sufficient to suppress the Zeeman energy and spin indices.

Solving the Schrödinger equation using the finite difference time domain method

We now consider our first example of a circularly symmetrical quantum dot confined in an harmonic potential with a magnetic field perpendicular to the 2D plane. Using the symmetric gauge $\mathbf{A} = (-\frac{1}{2}By, \frac{1}{2}Bx, 0)$, the Hamiltonian of Eq. (3) is

$$\hat{H} = -\frac{\hbar^2}{2m}\left[\frac{\partial^2}{\partial x^2} + \frac{\partial^2}{\partial y^2}\right] + i\hbar\frac{\omega_C}{2}\left[y\frac{\partial}{\partial x} - x\frac{\partial}{\partial y}\right] + \frac{m}{2}(\frac{\omega_C^2}{4} + \omega_0^2)(x^2 + y^2) \qquad (4)$$

where $\omega_0$ is the angular frequency of the confining harmonic oscillator potential and $\omega_C = eB/m$ is the cyclotron angular frequency. The energy eigenstates of this system are well known [2-4]. The eigenvalues are real and given by

$$E_{nl} = (2n + |l| + 1)\hbar(\frac{1}{4}\omega_C^2 + \omega_0^2)^{1/2} - \frac{1}{2}l\hbar\omega_C \qquad (5)$$

$n = 0,1,2,\cdots$ and $l = 0,\pm 1,\pm 2\cdots$ are the radial and azimuthal quantum number respectively. The complex eigenfunctions are

$$\psi_{nl}(r,\theta) = \left[\frac{b}{2\pi l_0^2}\frac{n!}{2^{|l|}(n+|l|)!}\right]^{\frac{1}{2}} \exp\left(-il\theta - \frac{br^2}{4l_0^2}\right)\left(\frac{\sqrt{b}r}{l_0}\right)^{|l|} L_n^{|l|}(\frac{br^2}{4l_0^2}) \qquad (6)$$

where $L_n^{|l|}$ are associated Laguerre polynomials, $b = \left(1 + 4(\omega_0/\omega_C)^2\right)^{\frac{1}{2}}$ and $l_0 = (\hbar/eB)^{\frac{1}{2}}$. We now show that eigenfunctions and eigenvalues of Eq. (4) can be accurately obtained using the FDTD method.

For numerical validation of the FDTD method we use parameters $\Delta x = \Delta y = 0.1$ and $\Delta\tau = \Delta x^2/6$ to define a grid. For computational convenience we confine the system in computational square of length 10. The numerical eigenvalues for the lowest four energy states ($\psi_{00}, \psi_{01}, \psi_{02}$ and $\psi_{0-1}$) as a function of magnetic field are shown in Fig. 1. It is seen that the numerical results with this specified grid size are in excellent agreement with the exact results Eq. (5). We found the errors of the FDTD numerical eigenvalues to be of order 0.1%. Comparisons of all numerical eigenfunctions with Eq. (6) also show excellent agreement. An example of this comparison is given in Fig. 2 for the eigenfunction $\psi_{03}$. We note that accuracy of the FDTD results can be improved, whenever required, by using smaller grid spacing and larger computational length.

Solving the Schrödinger equation using the finite difference time domain method

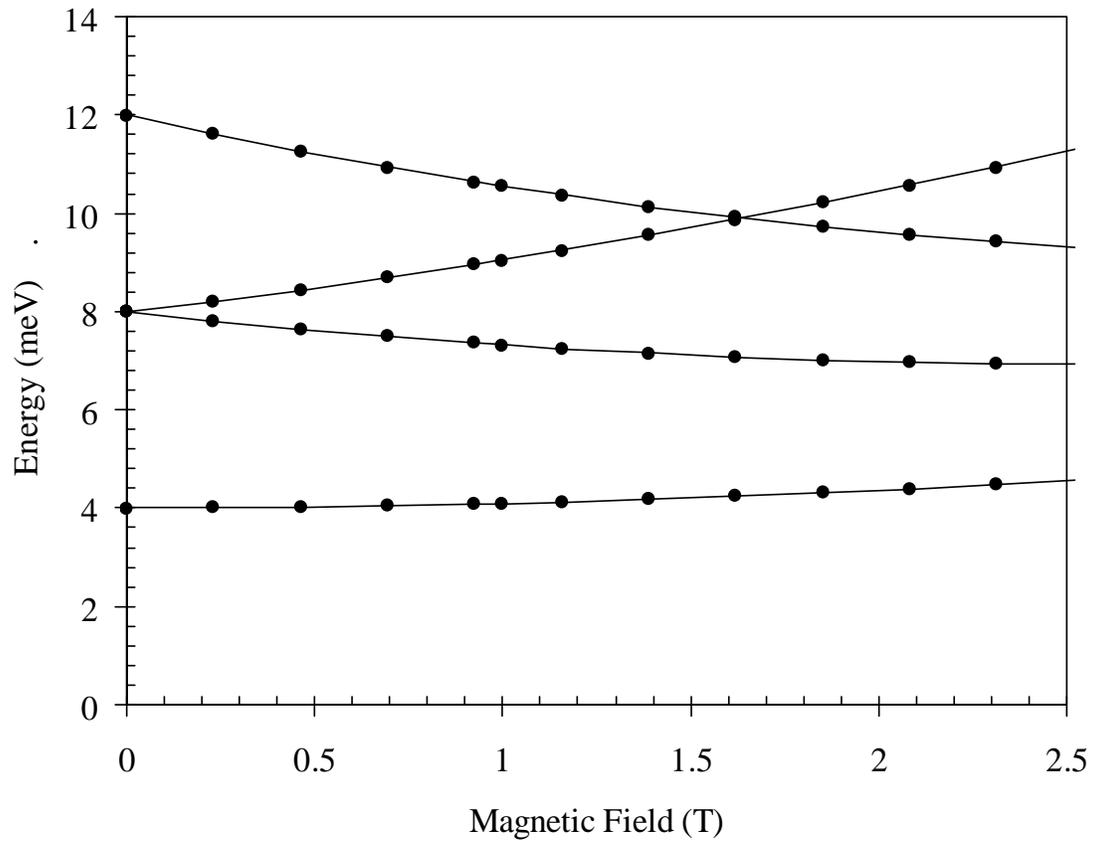

Figure 1. Comparison of numerical results for the lowest four energy eigenvalues (dots) and exact analytical results (lines) for an electron confined in 2D harmonic oscillator potential in a perpendicular magnetic field. Parameters used are appropriate for a quantum dot in GaAs, $\hbar\omega_0 = 4 \text{ meV}$ and a ratio of effective electron mass m to bare electron mass of 0.067.

Solving the Schrödinger equation using the finite difference time domain method

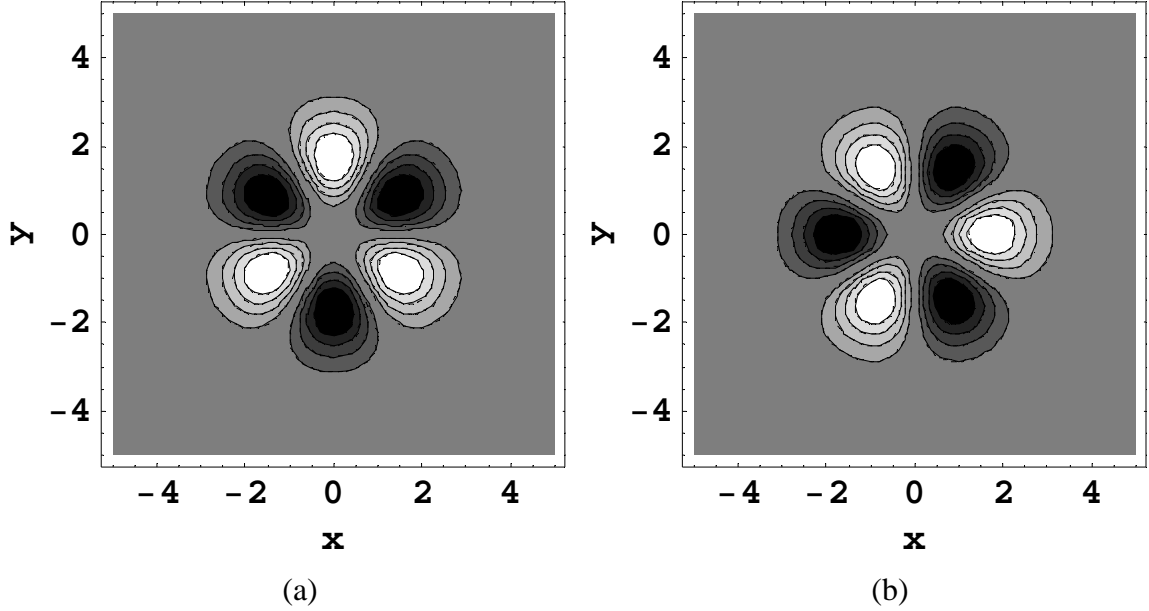

(a)　　　　　　　　　　　　　　(b)

Figure 2. Numerical results for (a) real and (b) imaginary part of the sixth excited state (i.e. $\psi_{03}$) for an electron confined in harmonic oscillator potential in a magnetic field $B = 1\,\text{T}$. The energy scale and effective electron mass ratio are as in Fig. 1. The contour lines with black shading have values starting from $-0.05$ (outer line) and with decrement of $0.05$. The contour lines with white shading have values starting from $0.05$ (outer line) and with increment of $0.05$. The exact contour lines coincide with the numerical contour lines on this scale so are not shown separately.

We next turn to our second example, an anisotropic 2D single electron quantum dot in an asymmetric quartic confining potential given by

$$V(x,y) = \frac{1}{2}m\omega_0^2(x^2 - y^2) + \frac{1}{2}(m^2\omega_0^3/\hbar)(x^4 + y^4) \tag{7}$$

As in the first example, the magnetic field is perpendicular to the plane and the symmetric gauge $\mathbf{A} = (-\frac{1}{2}By, \frac{1}{2}Bx, 0)$ is used. We again use parameters $\Delta x = \Delta y = 0.2$ and $\Delta \tau = \Delta x^2/6$ to define a grid, and the system is confined in computational square of length 10.

Since no exact analytical results are available for this confining potential, we also computed numerically the eigenvalues and eigenvectors by using a standard diagonalization method. This permits a direct consistency check on the accuracy. The comparison of energy eigenvalues using these two numerical methods is shown in Fig. 3 and it is seen that the two methods are in excellent agreement. We found that energy eigenvalues computed by the two methods agree to three part in 10,000. The FDTD results for energy eigenfunctions are also in agreement with those of the diagonalization method. An example of this is shown in Fig. 4.

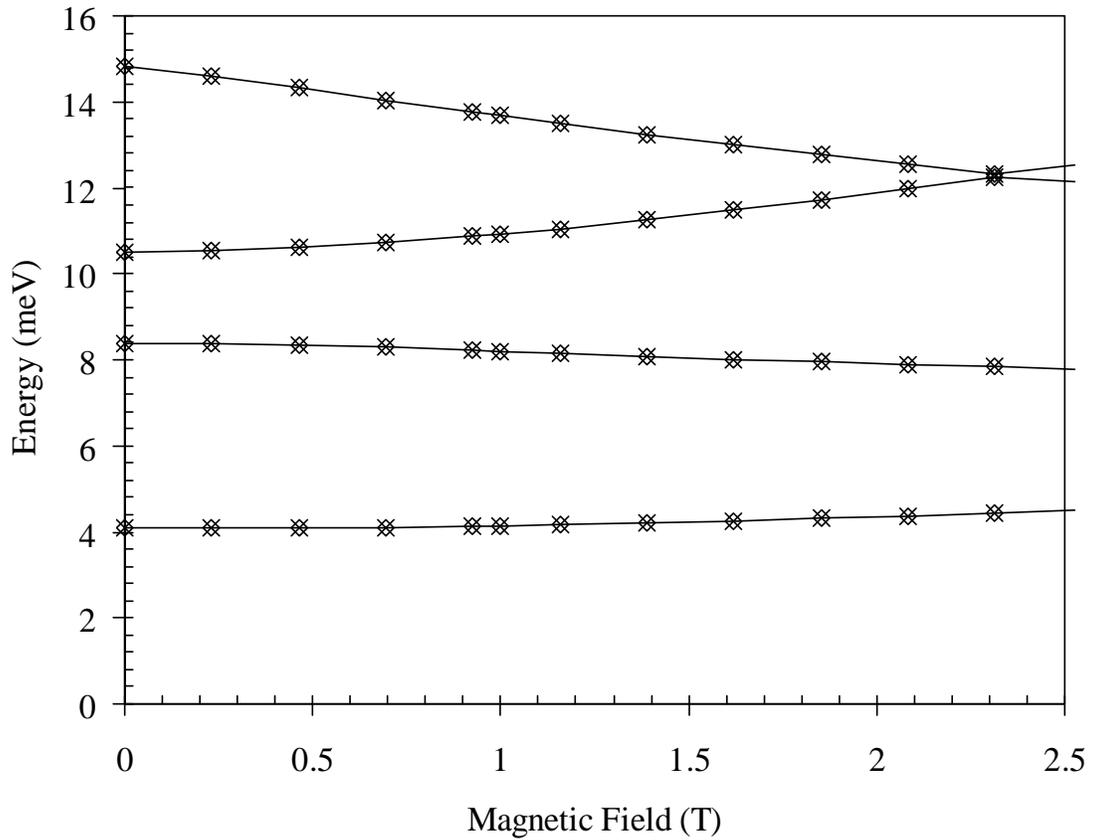

Figure 3. Numerical results for the lowest four energy states (crosses is for FDTD method and diamonds is for diagonalization method) for an electron confined in a quartic potential (Eq. 7) in a perpendicular magnetic field. The energy scale $\hbar\omega_0 = 4 \text{ meV}$ and effective electron mass ratio of 0.067 are as in Fig. 1. The numerical results obtained by matrix diagonalization coincide with the FDTD results on this scale and the line is a guide to the eye. The numerical differences between the two methods are found to be less than $3\times10^{-4}$ for all points.

Solving the Schrödinger equation using the finite difference time domain method

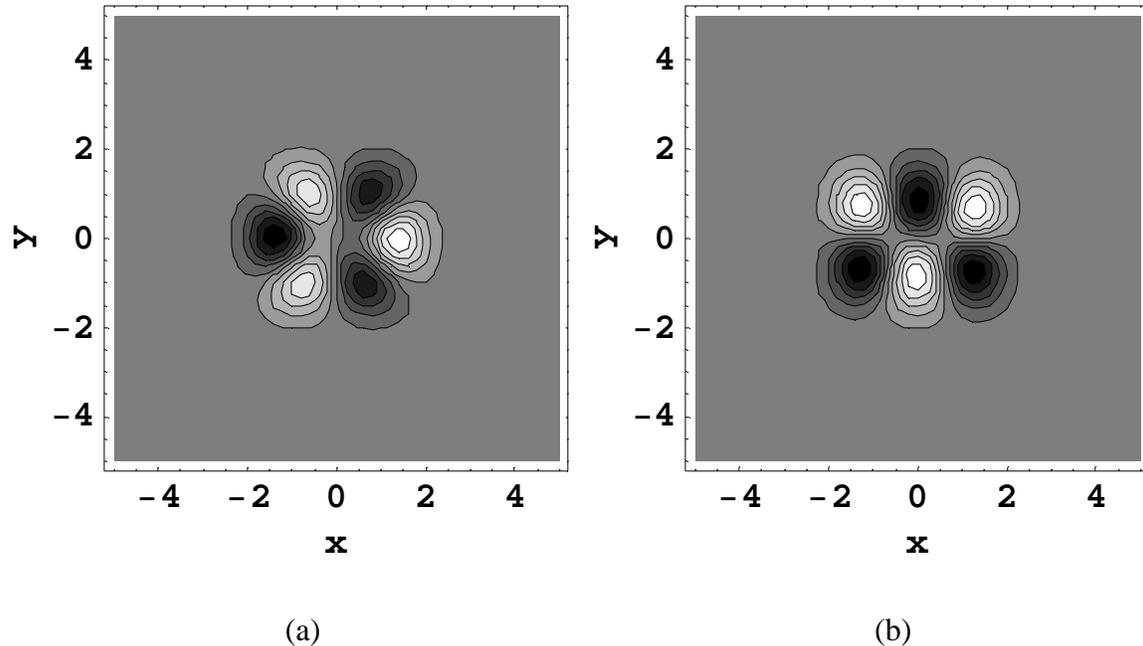

(a) (b)

Figure 4. The FDTD results for (a) real and (b) imaginary part of the sixth excited state for an electron confined in a quartic potential in magnetic field $B = 1\,\text{T}$. The energy scale and effective electron mass ratio are as in Fig. 1. The contour lines with black shading have values starting from $-0.05$ (outer line) and with decrement of $0.05$. The contour lines with white shading have values starting from $0.05$ (outer line) and with increment of $0.05$. The results of the diagonalization method coincide with the FDTD numerical contour lines so they are not shown separately.

We have also applied the FDTD method to a number of other single electron systems in 2D confining potentials, with no particular symmetries, in an external magnetic field. In all cases, convergence of eigenvalues and eigenfunctions was good and the specified accuracy requirements were easily obtainable by appropriate choice of computational grid.

**Conclusion**

We have considered the extension of our FDTD method [1] to cases where the Hamiltonian is Hermitian but has explicit factors of $i$ so that eigenfunctions are complex. This is the case when external magnetic fields are applied and time reversal symmetry is broken. Examples of an electron in a confining potential in a 2D plane, with a constant magnetic field perpendicular to the plane, were given to illustrate numerical applications of the procedure. The accuracy of computed eigenvalues and eigenfunctions was excellent. We emphasize that this FDTD method can be easily applied for general potentials without any particular symmetries. We conclude that the FDTD algorithm is an accurate and powerful method for numerical solution of the Schrödinger equation for a single electron system in a general confining potential in an applied magnetic field. It is

Solving the Schrödinger equation using the finite difference time domain method expected that this method will be particularly useful for determining the finite temperature properties of a range of low symmetry quantum systems.

Acknowledgements

Part of this work was done while one of the authors (DJWG) was attending a workshop on many-body theory at Centro De Giorgi, Pisa, Italy. The hospitality of the Centro De Giorgi and discussions with N.H. March are acknowledged.